\documentclass[journal, 10pt, twocolumn]{IEEEtran}

\usepackage[margin=1in]{geometry}
\usepackage{url}

\usepackage{amsmath}
\usepackage{amsfonts}

\usepackage{algpseudocode}
\usepackage{algorithm}

\usepackage{graphicx}
\usepackage{epstopdf}
\usepackage{subfigure}

\graphicspath{{figs/}}

\newcommand{\Xc}{\mathcal{X}}
\newcommand{\Sc}{\mathcal{S}}

\newcommand{\Rc}{\mathcal{R}}

\newcommand{\Pc}{\mathcal{P}}
\newtheorem{theorem}{Theorem}[section]

\newtheorem{lemma}[theorem]{Lemma}

\title{Compensating Demand Response Participants Via Their Shapley Values}
\author{Gear\'{o}id O'Brien, Abbas El Gamal and Ram Rajagopal
\thanks{G. O'Brien and A. El Gamal are with the Department of Electrical Engineering,  Stanford University (e-mail: gobrien@stanford.edu,abbas@ee.stanford.edu).  G. O'Brien is supported by a Stanford Graduate Fellowship and El Gamal is partially supported by the TomKat Center.}
\thanks{R. Rajagopal is with the Department of Civil and Environmental Engineering, Stanford University (e-mail: ramr@stanford.edu). Rajagopal is supported by a Powell Foundation Fellowship and the TomKat Center.}
}

\date{}

\begin{document}
\bstctlcite{IEEEexample:BSTcontrol}
\maketitle

\begin{abstract}
Designing fair compensation mechanisms for demand response (DR) is challenging. This paper models the problem in a game theoretic setting and designs a 
payment distribution mechanism based on the Shapley Value. As exact computation of the Shapley Value is in general intractable, we propose estimating it using a reinforcement learning algorithm that approximates optimal stratified sampling. We apply this algorithm to two DR programs that utilize the Shapley Value for payments and quantify the accuracy of the resulting estimates. 
\end{abstract}

\section{Introduction}
\label{sec:introduction} 

Demand response, or adjusting the aggregate load profile as a means to help balance supply and demand in electricity grids, is becoming an important approach to improve grid reliability. Utilities such as PG\&E~\cite{pgande_dr} allow for third party operators to administer DR programs through schemes such as \emph{Aggregator Managed Portfolios}.  These operators (or `aggregators') are responsible for most of the program details, including marketing, enrollment, and payments to participants.  

Designing, implementing, and operating large scale DR programs is a non-trivial task, however.  Ensuring that participants remain enrolled in the scheme --- and that it is also appealing to new participants --- relies in part on a fair and attractive compensation mechanism.  The typical mechanism design problem focuses on distributing the total revenue accrued by an aggregator to all participants, in proportion to their contribution in addition to achieving fairness and enrollment incentives. 

Design of demand response schemes has been extensively investigated in the literature \cite{Borenstein2002}.   Several papers have investigated price based mechanisms for various types of response capabilities \cite{Li2011,Gatsis2011, Conejo2010, ONeill2010, Dong2012, Kim2011}. Mechanisms based on cooperative games have been investigated more recently \cite{Zhu2012, Saad2012}, but attention has focused on specific formats where consumers make choices under parameterized utility functions. Instead in this paper, we propose a simple payment scheme based on a traditional cooperative game solution concept: the Shapley Value. 

Understanding the advantage of using the Shapley Value when compared to other conventional distribution methods is crucial to assessing the contribution of this paper.  In the proposed mechanism each consumer that agrees to participate in a DR program receives a payment for cooperating and forming a coalition. The \emph{value} of a coalition is a function measuring how close that coalition comes to achieving the goal of the DR program. A simple such function encapsulates the \emph{penalty} (in dollars) imposed by the utility on the aggregator for failing to meet an agreed upon commitment for the chosen coalition.  The aggregator needs to decide then how to distribute the penalty fairly among the participants.  Distributing penalties or payments in DR schemes is integral to their success, primarily in situations where participants are free to choose from a number of schemes.  The \emph{Shapley Value} solution concept provides a fair and unique method for distributing the total penalty when the penalty function satisfies some conditions.  Although the use of the word `fair' may seem vague in this context, it is a precisely defined term satisfying the following four concepts, which reasonable distribution schemes should satisfy:
\begin{itemize}
	\item[] \textbf{Efficiency: } The entire payment or penalty is divided among the participants (no excess remains).
	\item[] \textbf{Symmetry: } Two participants that contribute equally are rewarded equally.
	\item[] \textbf{Null Player: } Participants that do not contribute receive no payoff.
	\item[] \textbf{Linearity: } The total payoff rewarded for contributing to two programs is the sum of the payoff that would have been awarded for contributing to each of the two programs individually.
\end{itemize}
Surprisingly, the Shapley Value can be proven to be the \emph{only} payment distribution method that satisfies the four axioms above, with the added benefit that the solution is unique. The issue of appropriately allocating penalties to participants is an important problem in demand response that has not been the focus of much research.  The Shapley Value is an attractive solution to the problem.

The Shapley Value has been previously used in studies on electrical energy generation and transmission.  In~\cite{junqueira2007aumann}, the Shapley Value is used to allocate transmission service costs among network users in energy markets.   In~\cite{baeyens2011wind} the aggregation of wind power producers is studied using coalition game theory and show that the resulting game is not convex so the Shapley Value may not be appropriate. 

The most challenging aspect in utilizing the Shapley Value is its computational intractability. For a DR program with $n$ participants, the value function must be evaluated $n\,2^n$ times. A modest DR program with $n=500$ requires $1.5 \times 10^{153}$ function evaluations. Approximation approaches have been proposed to mitigate this problem. They rely on simple schemes to selectively perform function evaluations. Shapley proposed a Monte-Carlo random sampling technique ~\cite{generatingShapleyMonteCarlo}, extended in ~\cite{Bachrach:2010:API:1713753.1713758}  and ~\cite{castro2009polynomial} to achieve desired accuracy levels in polynomial time.  Such mechanisms neither exploit relevant properties of the value function nor enforce important constraints such as budget balance. 

This paper, which is an extended and more complete version of~\cite{O'Brien:2013:ECS:2487166.2487208}, proposes a Shapley Value based distribution of Demand Response payments (or penalties) and a new algorithm for estimation that is significantly faster and more accurate than prior approaches. The rest of this paper is organized as follows.  Section~\ref{sec:intro_shapley_value} provides a brief introduction to the Shapley Value.  Section~\ref{sec:estimating_shapley} describes a method of estimating the Shapley Value using a pseudo-random sampling technique that significantly reduces the variance of the estimate when compared to random sampling.  In Section~\ref{sec:DR_programs}, we analyze two simple demand response programs and utilize the Shapley Value as a means of compensating the participants in the scheme.


\section{Demand Response and the Shapley Value}
\label{sec:intro_shapley_value}

Consider a set $\Xc = \{1, 2, \ldots, n\}$ of $n$ participants in a DR scheme.  A ``participant'' could be a user or a user-load if such granularity is available.  For the set $\Sc \subseteq \Xc$, define the value function (or characteristic function) $v(\Sc)$ as the total penalty imposed on the participants in $\Sc$ if they do not achieve the DR goal.  For example, if the goal of the DR program is to provide reserve by reducing load levels, the value function may be taken as
\begin{align}
v(\Sc) = -\left[ \sum_{i \in \Sc} \left( X_i - \tilde{X}_i\right)\right]_+,
\label{eqn:example_vS}
\end{align}
where $X_i \in \mathbb{R}$ is the amount participant $i$ agreed to reduce its load by and $\tilde{X}_i \in \mathbb{R}$ is the amount it actually reduced it by.  As $\left[x \right]_+ = \max\{x,0\}, \quad v(\Sc)$ is non-zero when the aggregate discrepancy is greater than $0$.  Additional choices of value functions are discussed in detail in Section~\ref{sec:DR_programs}. We assume the general function $v(\Sc)$ is \emph{submodular} (see \cite{TheValueofanNPersonGame}  for a definition). 

The operator wishes to distribute the total penalty $v(\Xc)$ among the $n$ participants in a fair manner, dependent on their relative contributions to the goal of the DR scheme.  We denote the penalty assigned to participant $i$ as $\phi_i$. Hence the total penalty is
\begin{align}
v(\Xc) = \sum_{i=1}^n \phi_i.
\label{eqn:budget_balanced}
\end{align}

Shapley proposed a solution to the distribution of the total penalty that is both unique and fair when penalty functions are submodular ~\cite{TheValueofanNPersonGame} .  For a given participant, it is the mean marginal contribution of that participant to all possible coalitions of the other participants.  Defining $\Xc_{-i}$ to be the set of all participants after removing participant $i$, the marginal contribution of participant $i$ to a coalition $\Sc$, $\Sc \subseteq \Xc_{-i}$, is 
\begin{align}
	\rho_i(\Sc) &= v(\Sc \cup \{i\}) - v(\Sc).\label{eqn:marginal_contribution}
\end{align}
Furthermore, we define $\Rc$ to be one of the $n!$ permutations of the participants in $\Xc$, and $\Pc_i^{\Rc}$ to be the ordered set preceding $i$ in $\Rc$.  The Shapley Value is defined as
\begin{align}
	\phi_i(v) &= \frac{1}{n!} \sum_{\Rc} \rho_i(\Pc_i^\Rc).\label{eqn:shapley_value_1}
\end{align}
Clearly, direct calculation of the Shapley Value using equation~\eqref{eqn:shapley_value_1} is intractable.  For example, with a value function as seemingly elementary as that shown in~\eqref{eqn:example_vS}, the function is of a form $\max\{x,0\}$ and the Shapley value is computationally intensive to calculate.  Hence, it is typically estimated using sampling techniques.  The following section details a computationally efficient method for estimating the Shapley Value via sampling.  

\noindent {\bf Remark}.  For brevity we suppress the $v$ and denote the Shapley Value by  $\phi_i$. The value function assumed should be clear from the context.

\subsection{Value Functions and Demand Response Schemes}
\label{ssec:value_functions_and_DR}
In order to use the Shapley value as a distribution mechanism of a DR scheme, the scheme itself must be representable as a value function.  This function, defined over subsets of the participants, returns the penalty that will be imposed on the operator of the DR scheme.  Value functions are particularly suited to schemes where loads are controllable to some degree.  The formulation of the value function is left to the designer of the DR scheme as this paper is neither involved with choosing the value function itself nor designing DR schemes.


\section{Estimating the Shapley Value}
\label{sec:estimating_shapley}

Before describing our algorithm for estimating the Shapley Value, we need the following alternative formulation.  Grouping the terms in equation~\eqref{eqn:shapley_value_1} in which the participants to the left of $i$ are the same gives the alternative form for the Shapley Value
\begin{align}
	\phi_i &= \sum_{\Sc \subseteq \Xc_{-i}} \frac{|\Sc|!(|\Xc|-|\Sc|-1)!}{n!} \rho_i(\Sc). \label{eqn:shapley_value_2}
	\intertext{Further grouping by the number of terms in $\Sc$, defining $j = |\Sc|$, and recalling that $n = |\Xc|$ we obtain}
	\phi_i &= \sum_{j = 0}^{n-1} \sum_{\Sc \subseteq \Xc_{-i}} \left(\frac{j!(n-j-1)!}{n!}\right)\rho_i(\Sc)\notag\\
	&= \frac{1}{n}\sum_{j=0}^{n-1} \sum_{\Sc \subseteq \Xc_{-i}}\left(\frac{(n-1)!}{j!(n-1-j)!}\right)^{-1}\rho_i(\Sc)\notag\\
	&=\frac{1}{n}\sum_{j=0}^{n-1} \sum_{\Sc \subseteq \Xc_{-i}} \binom{n-1}{j}^{-1} \rho_i(\Sc).\notag
\end{align}
The inner sum can be considered as an expectation over a uniform probability mass function, hence we can write	
\begin{align}
	\phi_i&=\frac{1}{n}\sum_{j=0}^{n-1}\mathbb{E}[\rho_i(\Sc)].
	\label{eqn:shapley_average_expected_value}
\end{align}
This form of the Shapley Value suggests an estimation approach based on stratified sampling~\cite{rao2000sampling}.  For participant $i \in \{1,2,\ldots,n\}$, let stratum $j$ be the set of marginal contributions of that participant to every subset $\Sc \subseteq \Xc_{-i}$ of size $|\Sc| = j$.  We randomly and independently draw $N^i_j$ samples $\rho^i_{1,j}, \ldots, \rho^i_{N^i_j,j}$ from each stratum $j$.  Define the sample mean for participant $i$ as the random variable
\begin{align}
T(\rho^i_{k,j}) &= \frac{1}{n}\sum_{j=0}^{n-1}\frac{1}{N^i_j}\sum_{k=0}^{N^i_j}\rho^i_{k,j}\notag\\
&=\frac{1}{n}\sum_{j=0}^{n-1}\overline{\rho}_j^i\label{eqn:T_statistic},
\end{align}
where the random variable $\overline{\rho}_j^i$ is the sample mean of the data drawn from stratum $j$. This sample mean is a linear unbiased estimate of $\phi_i$ and would be a reasonable estimate of $\phi_i$ except for the fact that the sum of the estimates may not be equal to the total budget $v(\Xc)$, which would violate the efficiency axiom of the Shapley Value. We therefore use the sample averages as basis for computing the maximum likelihood (ML) estimates of the Shapley Values with the budget constraint as follows.

Assume that the number of samples from each stratum is sufficiently large so that we can use the central limit theorem to approximate the distribution of $\overline{\rho}_j^i$ by a Gaussian with mean $\mu_{j,i} = \mathbb{E}[\rho_i(\Sc)]$ and variance $\sigma_{j,i}^2$. By independence of the sample averages, it follows that the variance of $T(\rho^i_{k,j})$, 
\[
\sigma_i^2 = \frac{1}{n^2}\sum_{j=0}^{n-1}\sigma_{j,i}^2,
\]
and $T(\rho^i_{k,j}) \sim \mathcal{N}(\phi_i, \sigma_i^2)$. 

By independence of the sample averages $f(T(\rho^i_{k,j}) | \phi_i)$, $i\in \Xc$, the likelihood function can be written as 
\begin{align}
f( T(\rho^1_{k,j}), \ldots, T(\rho^n_{k,j}) |  \phi_1 \ldots \phi_n) &= \prod_{i=1}^n f\left(T(\rho^i_{k,j}) | \phi_i\right).\label{eqn:likely_prod}
\end{align}
Since the sample averages $T(\rho^i_{k,j})$ are Gaussian, we consider the log likelihood function
\begin{align}
\sum_{i=1}^n\log\left( f\left(T(\rho^i_{k,j}) | \phi_i\right)\right)&= \zeta - \sum_{i=1}^n\frac{\left(T(\rho^i_{k,j}) - \phi_i\right)^2}{2\sigma^2_i},
\end{align}
where $\zeta$ is not a function of $\phi_i$. To obtain the ML estimates of the Shapley Values we then need to solve the optimization problem with respect to $\phi_i$:
\begin{align}\label{eqn:opt}
\underset{\phi_i}{\textrm{maximize}}\enspace &\zeta - \sum_{i=1}^n\frac{\left(T(\rho^i_{k,j}) - \phi_i\right)^2}{2\sigma^2_i} \nonumber\\
\textrm{subject to}\enspace & \sum_{i=1}^n \phi_i = v(\Xc).
\end{align}

To solve this problem, we form the Lagrangian 
\begin{align*}
\mathcal{L} = \zeta - \sum_{i=1}^n\frac{\left(T(\rho^i_{k,j}) - \phi_i\right)^2}{2\sigma^2_i} + \lambda\left( v(\Xc) - \sum_{i=1}^n\phi_i\right).
\end{align*}
This is a convex optimization problem and has a simple analytical solution.

\begin{theorem}
The ML estimates of $\phi_i$ are given by
\begin{align}
\hat{\phi}_i = T(\rho_{k,j}^i) - \frac{\sigma_i^2}{\sum_{m=0}^n\sigma_m^2}\big( \hat{v}(\Xc) - v(\Xc)\big),
\label{eqn:MLE}
\end{align}
where $\hat{v}(\Xc) = \sum_{i=1}^n T(\rho_{k,j}^i)$.  
\end{theorem}
\smallskip

Note that all properties of the Shapley Value (efficiency, symmetry, null player, linearity~\cite{Owen_GameTheory}) hold in expectation in equation~\eqref{eqn:MLE}, with the added benefit that the budget is \emph{always} balanced, as the constraint in the optimization problem~\eqref{eqn:opt} is satisfied.

\subsection{Sample allocation}

We now turn our attention to the question of how many samples we should select from each stratum. Suppose we have a total budget of $N$ samples per participant, i.e., $\sum_{j=0}^{n-1} N_j^i = N$ for every $i \in \Xc$. How do we divide them among the strata? One reasonable approach would be to allocate the samples for each participant $i$ to minimize the variance of the sample mean $T(\rho_{k,j}^i)$ subject to $\sum_{j=0}^{n-1}N_j^i = N$.  The following shows that the optimal sample allocation is the Neyman allocation~\cite{rao2000sampling} for equal weighting.

\begin{lemma}
\label{lma:neyman_allocation}
The minimum variance of $T(\rho_{k,j}^i)$ subject to $\sum_{j=0}^{n-1}N_j^i = N$ is
\begin{align}
\sigma^2_{i, \mathrm{SD}}&= \frac{1}{N} \mathrm{mean}(\sigma_{j,i})^2,\label{eqn:var_SD}
\end{align}
where $\sigma_{j,i}$ is the standard deviation of the population in stratum $j$ for participant $i$.  The value of $\mathrm{mean}(\sigma_{j,i})$ is calculated by averaging over the $n$ values of $\sigma_{j,i}$ for participant $i$. 

The values of $N_j^i$ that achieve this minimum are
\begin{align*}
N_j^i = \frac{\sigma_{j,i}}{\sum_{m=0}^{n-1}\sigma_{m,i}},\; j \in \{0,1,\ldots,n-1\}.
\end{align*}
\end{lemma}
The proof of this lemma is given in the Appendix~\ref{app:A}.

It is interesting to compare the achievable variance of the sample means using the above optimal stratified sampling to the more commonly used uniform sampling.  With uniform sampling, we draw $N$ samples independently at random from the set of marginal contributions of participant $i$ without taking strata into consideration. The variance of the sample average for this approach is
\begin{align}
 \sigma^2_{i, \mathrm{RS}} &= \frac{1}{N}\left[ \mathrm{mean}(\sigma_{j,i}^2) + \mathrm{var}(\mu_{j,i})\right],\label{eqn:var_RS}
\end{align}
where $\mu_{j,i}$ is the mean value of the population in stratum $j$ for participant $i$.  The value $\mathrm{var}(\mu_{j,i})$ is calculated as the variance of the $n$ values of $\mu_{j,i}$ for participant $i$.  The proof of this fact is given in Appendix~\ref{app:B}.  

Sampling according to the Neyman Allocation (lemma~\ref{lma:neyman_allocation}) requires prior knowledge of the standard deviation of each stratum for each participant, which is not realistic. A more practical approach would be to sample equally from each of the $n$ stratum, i.e., $N_j^i = N/n$.  With this allocation, the variance of the sample average is 
\begin{align}
\sigma^2_{i, \mathrm{ES}} &= \frac{1}{N}\mathrm{mean}(\sigma_{j,i}^2).\label{eqn:var_ES}
\end{align}
The proof of this fact follows readily from the proof of~\eqref{eqn:var_RS} and is omitted.  Comparing the variances for these three allocation strategies, we can clearly see that
\begin{equation}
\sigma^2_{i, \mathrm{SD}} \leq \sigma^2_{i, \mathrm{ES}} \leq \sigma^2_{i, \mathrm{RS}}.\label{eqn:var-comparison}
\end{equation}
Hence, it is always better to sample in proportion to standard deviations. In the following section we describe a reinforcement learning algorithm for estimating these standard deviations during sampling.

\subsection{Approximating Optimum Stratified Sampling}
\label{ssec:reinforcement_learning_alg}
Implementing an approximation to SD sampling is a typical reinforcement learning problem in which the algorithm seeks to exploit the information it has about the standard deviations of the strata to sample correctly, but must at the same time explore in order to accurately calculate these very standard deviations.  In our setting, the goal is to sample a specific (but unknown) number of times from each stratum.  This differs from the usual reinforcement learning problems where the goal is to converge on a single optimum action that maximizes the total reward.  This contrast means that some techniques (such as \emph{$\epsilon$-greedy}, \emph{Pursuit} and \emph{Reinforcement Comparison}) are not suitable, and other approaches must be altered to make them suitable for the problem at hand; see~\cite{sutton1998reinforcement} for information on reinforcement learning.
By comparison, stochastic methods~\cite{sutton1998reinforcement} which assign a probability to each action in accordance with the expected reward (or standard deviation in this case) are quite suitable to our setting.  

Our proposed algorithm~\ref{alg:approxSDsample} explicitly ``explores'' the problem space initially before gradually moving to an ``exploit'' phase in which it uses the results of the exploration to improve the sampling allocations. For participant $i$, the probability of sampling from stratum $j$ at sample $t \leq N$ is 
\begin{align}
\pi_{j,i}(t) = \epsilon(t) \frac{1}{n} + (1-\epsilon(t)) \frac{\hat{ \sigma}_{j,i}}{\sum_{m=0}^{n-1} \hat{ \sigma}_{m,i} },
\end{align}
\noindent
where $\hat{\sigma}_{j,i}$ is the current estimate of the standard deviation of stratum $j$.  
The choice of $\epsilon(t)$ is left to the user, but should be a decreasing function of $t$ with $\epsilon(0)=1$.  We implemented a number of such functions (including the stepped function described in~\cite{etore2010adaptive}) and found the most accurate to be the double sigmoid function   
\begin{align}
\epsilon(t) = \kappa - \frac{1}{1+e^{-\frac{t-\gamma N}{\beta N}}},
\label{eqn:epsilon_double_sigmoid}
\end{align}
where $\kappa$ is chosen to ensure $\epsilon(0) = 1$. Increasing $\gamma$ in the above equation reduces the percentage of samples used for exploration, and increasing $\beta$ increases the transition time from exploration to exploitation.  

\begin{algorithm*}[ht!]
\begin{algorithmic}
\Procedure{StandardDeviationSampling}{$N$, $i$}
\State $t\gets 1$	
\State $\hat{ \mu}_{j,i} \gets \underline{0}$ \Comment Estimate of $ \mu \in \mathbb{R}^n$
\State $\hat{ \sigma}_{j,i} \gets \underline{0}$ \Comment Estimate of $ \sigma \in \mathbb{R}^n$
\State $ c \gets \underline{0}$ \Comment Vector where $ c_j$ is the number of samples taken from stratum $j$, $ c \in \mathbb{R}^n$
\State $ m2 \gets \underline{0}$ \Comment Vector of the sum of squared differences from the current mean of stratum $j$, $ m2 \in \mathbb{R}^n$

\While {$t\leq N$}
	\State $\pi_{j,i}(t) \gets \epsilon(t) \frac{1}{n} + \left( 1 - \epsilon(t)\right) \frac{\hat{ \sigma}_j}{\sum_{m=0}^{n-1}  \sigma_{m,i}}$
	\State Choose stratum $j$ at random, weighted by $\pi_{j,i}(t)$.
	\State Choose a random coalition, $\Sc \subseteq \Xc_{-i}$ where $|\Sc| = j$.
	\State $x \gets \rho_i(\Sc)$ \Comment $x$ is a sample from stratum $j$
	\State $ c_j \gets  c_j + 1$ \Comment Update the count for stratum $j$
	
	\State $\Delta \gets x - \hat{ \mu}_{j,i}$
	
	\State $\hat{ \mu}_{j,i} \gets \hat{ \mu}_{j,i} + \frac{\Delta}{ c_j}$ \Comment Online update for estimate of $ \mu_{j,i}$
	
	\State $ m2_j \gets  m2_j + \Delta (x - \hat{ \mu}_{j,i})$
	\State $\hat{ \sigma}_{j,i} \gets \sqrt{\frac{ m2_j}{ c_j - 1}}$ \Comment Online update for estimate of $ \sigma_{j,i}$
\EndWhile

\State $T(\rho_{j,k}^i) \gets \mathrm{mean}(\hat{ \mu_{j,i}})$
\State $\sigma_i^2 = \frac{1}{n^2}\sum_{j=0}^{n-1} \hat{\sigma}_{j,i}^2$
\State \Return $T(\rho_{j,k}^i)$, $\sigma_i^2$
		
\EndProcedure
\end{algorithmic}
\caption{Approximating SD sampling}
\label{alg:approxSDsample}
\end{algorithm*}

At each step $t$, Algorithm~\ref{alg:approxSDsample} chooses stratum $j$ with probability $\pi_{j,i}(t)$ for participant $i$.  The probabilities are then updated for the next iteration.  The vector of standard deviations is updated in each step using a numerically stable algorithm from~\cite{Knuth:1997:ACP:270146}.  The algorithm returns the sample mean $T(\rho_{k,j}^i)$ for participant $i$ as well as the variance of that statistic, $\sigma_i^2$.  Once this has been calculated for all $n$ participants, the MLE can be computed using equation~\eqref{eqn:MLE} to ensure that the budget is balanced. 

If $\sigma^2_{i, \mathrm{SD}} << \sigma^2_{i, \mathrm{ES}}$, then implementing Algorithm~\ref{alg:approxSDsample} will significantly reduce the variance of the sample mean.  If however, $\sigma^2_{i, \mathrm{SD}} \approx \sigma^2_{i, \mathrm{ES}}$, then the benefit of the algorithm may well be outweighed by the complexity involved in the implementation and time involved in its execution.  Comparing $\sigma^2_{i, \mathrm{SD}}$ to $\sigma^2_{i, \mathrm{ES}}$, we have
\begin{align}
\frac{\sigma^2_{i, \mathrm{SD}}}{\sigma^2_{i, \mathrm{ES}}} = 1 + \frac{\text{var}(\sigma_{j,i}^2)}{\text{mean}(\sigma_{j,i})^2}.
\label{eqn:benefit_of_alg}
\end{align}
Hence, if $\text{var}(\sigma_{j,i}^2)/\text{mean}(\sigma_{j,i})^2 \approx 0$, sampling equally from each strata would be preferable.

\section{Demand Response Programs}
\label{sec:DR_programs}
We demonstrate the use of the Shapley Value to distribute compensation among the participants in two illustrative DR programs.  In practice, more complex value functions could be used achieving the similar gains in performance. 

\subsection{DR providing Reserve}
\label{ssec:DR_reserve_example}
In this DR example, each participant in the program agrees when requested to reduce its load by a predefined amount $X_i \in \mathbb{R}_+$ (loads can be reduced for example by dimming lights or controlling HVAC in a building).  The operator then offers a quantity $M$ of ``spinning reserve" to a utility where $\sum_{i=1}^n X_i \geq M, M \in \mathbb{R}_+$.  This ensures that the program has a leeway of $\Delta M = \sum_{i=1}^n X_i - M$.  When a demand response event is requested, each participant responds appropriately.  There may be a discrepancy between a participant's actual reduction in consumption, $\tilde{X}_i$, and the promised reduction amount $X_i$, which we denote $\Delta X_i = X_i - \tilde{X}_i$.  The value $\Delta X_i$ can be thought of as participant $i$'s contribution to the penalty which will be imposed if the total discrepancy across all $n$ participants exceeds $\Delta M$. We then take the value function for this DR program to be
\begin{align}
v(\Sc) = -q\left[\sum_{i \in \Sc} \Delta X_i - \Delta M \right]_+,
\label{eqn:char_function_DR_reserve}
\end{align}
where $\Sc \subseteq \Xc$ and $[x]_+ = \max \{x,0\}$ (i.e., the value function is non-zero only when the DR program is unable to meet the agreed upon reduction of $M$) and $q>0$ is a constant that converts energy to a penalty levied by the utility on the DR operator, which without loss of generality can be set to $1$.  Note that the value function in~\eqref{eqn:char_function_DR_reserve} is a form of a budget-additive function.  It is submodular and as such is compatible with the Shapley Value.

A value function such as that in~\eqref{eqn:char_function_DR_reserve} models well the Aggregator Managed Portfolio (AMP) Demand Response programs offered through PG\&E and operated by various third parties.  Non-compliance penalties are imposed on the aggregators: ``\emph{The aggregators are penalized if they fail to deliver their committed load reductions. The penalties vary based on the shortfall, with larger penalties for larger shortfalls. Aggregators determine compensation and/or penalties for their participating customers.}''  Using the Shapley Value as a means of determining compensation is of great relevance to such aggregators.

To compare the performance of the sampling techniques we discussed in Section~\ref{sec:estimating_shapley}, we consider a small set of $n=20$ participants so that we can compute the exact Shapley Value (ground truth).  

\begin{figure}[ht!]
\begin{center}%
\includegraphics[width=\columnwidth]{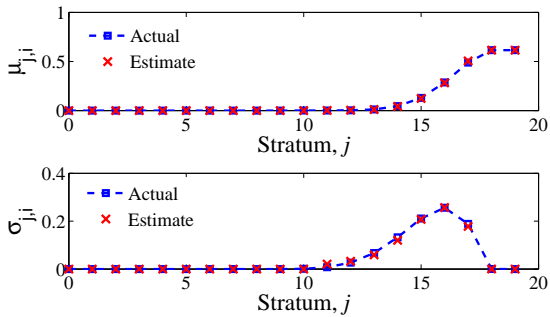}%
\caption{Mean (top) and standard deviation (bottom) for each stratum.  A red ``x'' indicates the final estimate from employing Algorithm~\ref{alg:approxSDsample} with $N=5000$ samples.}%
\label{fig:Mean_Std_DR_reserve}%
\end{center}%
\end{figure}

Figure~\ref{fig:Mean_Std_DR_reserve} plots the sample mean and standard deviation for each stratum when using the value function~\eqref{eqn:char_function_DR_reserve} for a representative participant $i$.  As can be seen, our stratified sampling algorithm which approximates sampling in proportion to the standard deviations shows significant improvements over both uniform and random sampling because strata 0 to 10 have zero mean (and standard deviation) and as such do not contribute to the Shapley Value and the samples taken from these strata in the uniform and random sampling methods are wasted.

\begin{figure}[ht!]%
\begin{center}%
\includegraphics[scale=0.45]{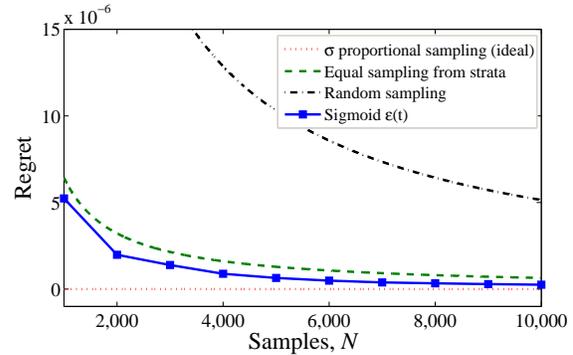}%
\caption{Decrease in `regret' as sample size increases for various $\epsilon(t)$ functions.  The upper dashed line is the regret from uniform sampling and the lower dashed line at zero regret corresponds to ideal sampling.}%
\label{fig:performance_of_algs}%
\end{center}%
\end{figure}

\noindent {\bf Remark}. As mentioned in Section~\ref{ssec:reinforcement_learning_alg}, we implemented the reinforcement learning algorithm using various $\epsilon(t)$ functions.  Figure~\ref{fig:performance_of_algs} plots the sample size against ``regret," defined as the difference between the variance of the Shapley Value estimate for a given $\epsilon(t)$ and that of the estimate calculated using exact SD sampling.  As can be seen, the sigmoid function (with $\gamma = 0.2$ and $\beta = 0.075$) which we use in all numerical results closely approximates ideal sampling.

\begin{figure}[ht!]%
\begin{center}%
\includegraphics[scale=0.45]{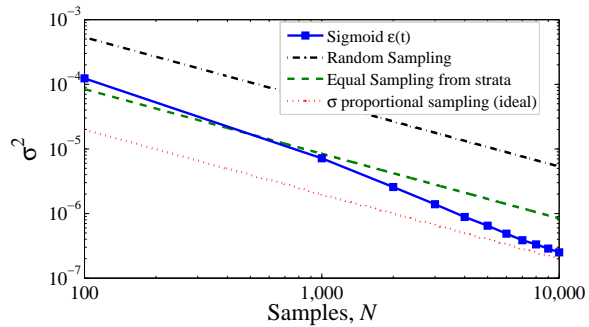}%
\caption{Change in variance as sample size increases.}%
\label{fig:Variance_SampleSize_DR1}%
\end{center}%
\end{figure}

Figure~\ref{fig:Variance_SampleSize_DR1} shows the reduction in variance of the Shapley Value estimate as we change our sampling technique, indicating that estimating the standard deviations in this scenario could significantly reduce error in the Shapley Value.  Implementing the proposed reinforcement learning algorithm reduces the empirical variance to a level approaching that achievable when the standard deviations are known in advance.

\begin{figure}[ht!]%
\begin{center}%
\includegraphics[scale=0.5]{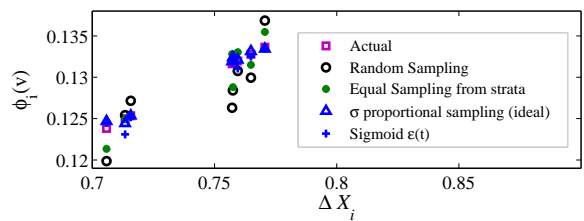}%
\caption{Actual and estimates of the Shapley Values for seven of the twenty participants in the DR program.  The estimates are calculated using $N=5000$.}%
\label{fig:all_shapley_values_reserve}%
\end{center}%
\end{figure}

In Figure~\ref{fig:all_shapley_values_reserve} we compare the estimated and actual Shapley Values for seven of the twenty participants for $0.7 \leq \Delta X_i \leq 0.8$. The estimates of the Shapley Values were calculated using  $N=5000$ samples.  For this single instance, it can be seen that employing stratified sampling reduces the error significantly when compared to random sampling.  

We determine the accuracy of the sampling methods using the mean squared prediction error
\begin{align}
\text{MSPE} = \mathbb{E}\left[\left(\phi_i - \hat{\phi}_i\right)^2\right].
\label{eqn:MSPE}
\end{align}
For ease of comparison, we normalize the MSPE for each method by the MSPE for $\sigma$ proportional sampling (the ideal method).  Table~\ref{tab:MSPE} contains the comparison results. It is clear that employing stratified sampling gives much better results than simple random sampling.  The learning algorithm significantly outperforms uniform stratified sampling and approaches the accuracy of ideal stratified sampling.

\begin{table}%
\caption{The normalized MSPE for the various sampling methods.}
\label{tab:MSPE}
\centering
\begin{tabular}{c||c}
Method & Normalized MSPE\\
\hline
Random Sampling & 26.3084\\
Equal Sampling from strata & 4.6521\\
$\sigma$ proportional sampling (ideal) & 1\\
Sigmoid $\epsilon(t)$ & 1.8050\\
\hline
\end{tabular}
\end{table}

\subsection{Deferrable Load following}
\begin{figure}[ht!]
\centering
\mbox{\subfigure{\includegraphics[width=0.485\columnwidth]{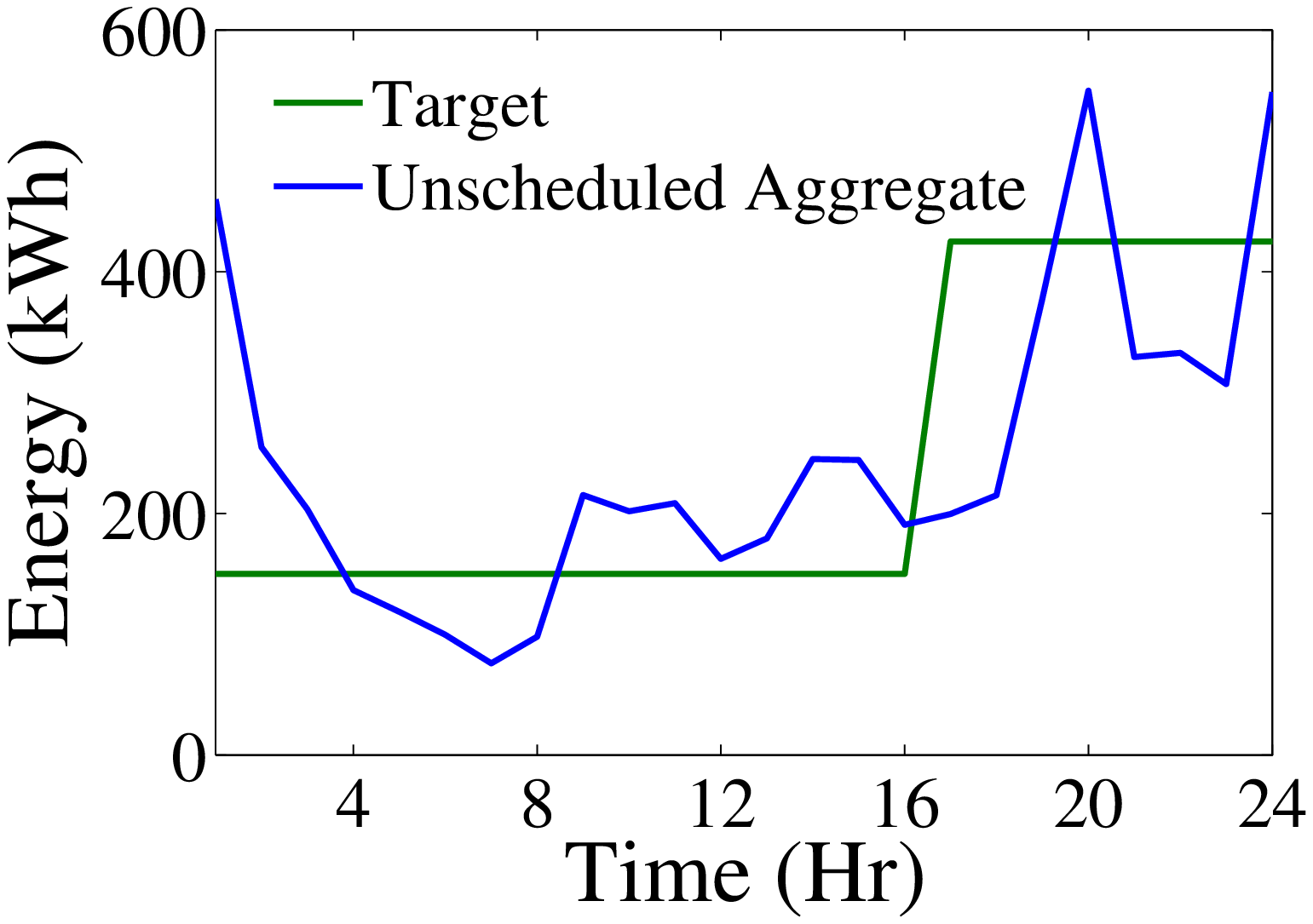}}
\subfigure{\includegraphics[width=0.485\columnwidth]{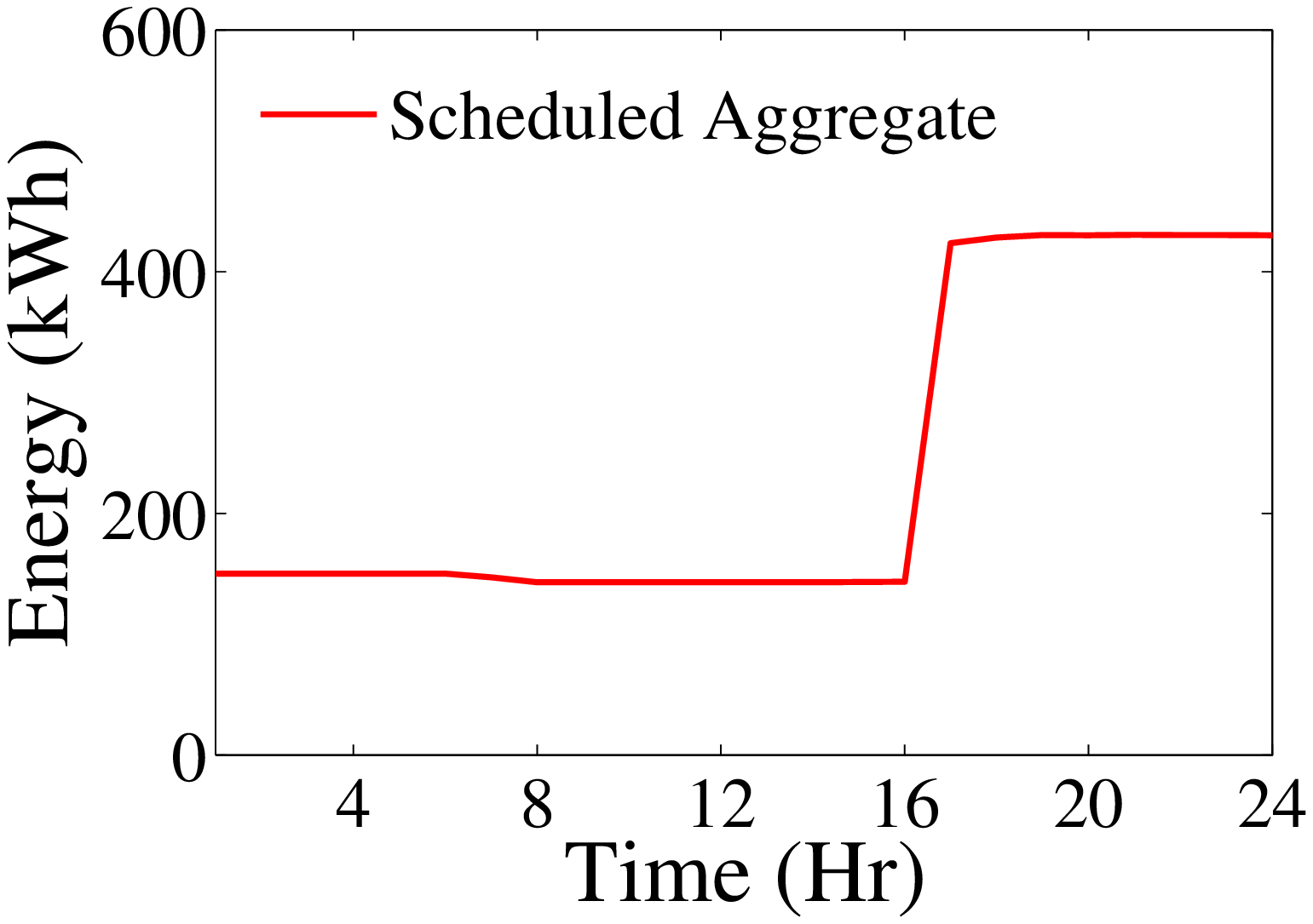} }}
\caption{Unscheduled aggregate on the left and scheduled aggregate on the right.} \label{fig:sample_scheduling}
\end{figure}

The second DR program we consider is deferrable load following, which is described in detail in~\cite{GOB_RR_deferrable_loads}.  In this program, $\boldsymbol X_i \in \mathbb{R}_+^T$ is a load profile for participant $i \in \Xc$ of $T$ time steps.  Given a set $\Sc \subseteq \Xc$, the operator wishes to \emph{schedule} (i.e., delay in time) a number of loads of the participants in $\Sc$ such that the new aggregate load profile $\textbf{s} \in \mathbb{R}_+^T$ approximates a predefined target load profile $\textbf{y} \in \mathbb{R}_+^T$.  This target profile is chosen in advance.  Each load has a maximum allowable delay which may be $0$ if the load is not schedulable.  This problem reduces to a ``knapsack packing'' exercise, hence it is NP hard.  To approximate a polynomial time solution, we use a greedy algorithm~\cite{GOB_RR_deferrable_loads} that analyses each load and its set of possible delays in order to maximize at each step the functional
\begin{align}
v(\Sc) = \frac{1}{T} \left( ||\textbf{y}||_2^2 - ||\textbf{y} - \textbf{s}||_2^2 \right),
\label{eqn:maximal_function}
\end{align}
where $\Sc \subseteq \Xc$, and $\textbf{s}$ is the aggregate load profile produced by appropriately scheduling the loads of the participants in $\Sc$.  The function $v(\Sc)$ is maximized when $\textbf{s} = \textbf{y}$.  The interested reader is referred to~\cite{GOB_RR_deferrable_loads} for a detailed discussion of this scheme.  In Figure~\ref{fig:sample_scheduling} we see the unscheduled aggregate together with the target profile $\textbf{y}$ on the left and the scheduled aggregate $\textbf{s}$ on the right. The data used to generate these figures and results was taken from the \emph{Plugwise} dataset, which contains plug level load data.  The raw data is divided into 24 hour blocks.  Each block is labeled with a user ID, timestamp, device description, and total energy consumption per hour for that device.

We assume that the operator of the DR program has realized revenue from operating the scheme and wishes to distribute this revenue fairly among the participants using the Shapley Value. 

The nature of the load following algorithm means that it is not very illuminating to analyze small load sets.  Hence in this case, we take a much larger load set containing $n=500$ load profiles.  For the load following scheme, each load profile has 24 data points corresponding to hourly smart-meter readings.  Again, a typical load was isolated from the Plugwise dataset and a detailed analysis was performed on that load. 

\begin{figure}[ht!]%
\begin{center}%
\includegraphics[width=\columnwidth]{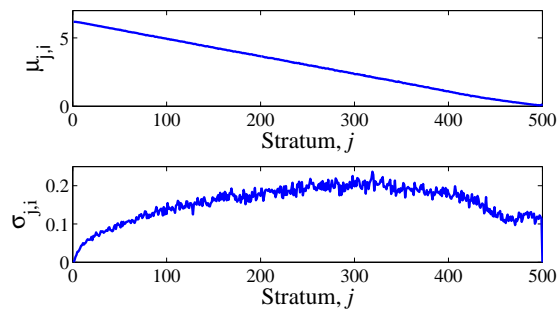}%
\end{center}%
\caption{Estimated mean (top) and estimated standard deviation (bottom) for each stratum.  Each stratum was sampled 200 times to estimate the mean and standard deviation.}%
\label{fig:Schedule_Mean_Std}%
\end{figure}

Figure~\ref{fig:Schedule_Mean_Std} plots the sample means and standard deviations of the strata when using the value function~\eqref{eqn:maximal_function}.  Calculating equation~\eqref{eqn:benefit_of_alg} for this instance results in $\sigma^2_{i, \mathrm{SD}} \approx \sigma^2_{i, \mathrm{ES}}$, and so the benefits of algorithm~\ref{alg:approxSDsample} will not be significant in this case.

\begin{figure}[ht!]%
\begin{center}%
\includegraphics[width=\columnwidth]{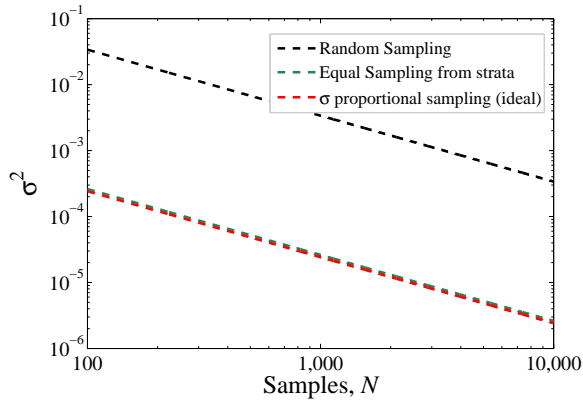}%
\end{center}%
\caption{Change in variance as sample size increases for the deferrable load scheduling program.  Note how close the curves are for both ideal sampling and uniform weighted sampling.}%
\label{fig:Variance_SampleSize_scheduling}%
\end{figure}

Using the estimates of the stratum means and standard deviations in figure~\ref{fig:Schedule_Mean_Std} to calculate the variance of the Shapley Value estimated using the three sampling techniques produces plots in figure~\ref{fig:Variance_SampleSize_scheduling}.  As the plots for change in variance using both ideal sampling and uniform weighted sampling are quite similar, the added benefit of using algorithm~\ref{alg:approxSDsample} will indeed be outweighed by the complexity involved in its implementation.  

\section{Conclusion}
\label{sec:conclusion}

This paper proposes the use of the Shapley Value to distribute the penalty among the participants in a DR program. As the Shapley Value is computationally intractable in general, we proposed a stratified sampling technique that reduces the number of samples needed to achieve a desired estimation accuracy while satisfying the budget balance constraint. We found that optimal stratified sampling requires prior knowledge of the standard deviations of the strata, which may not be available. As such, we proposed a reinforcement learning heuristic which estimates the standard deviations and uses them to adjust the sample allocation among the strata. We demonstrated the use of the Shapley Value in DR programs numerically, describing one scenario (DR providing reserve) where the reinforcement learning algorithm can significantly reduce the variance of the estimate and another scenario (deferrable load following) where sampling equally from each stratum is very nearly as effective as implementing the algorithm.

It should be noted also that this method is agnostic to the specifics of the characteristic function under analysis, and can therefore be used to estimate the Shapley Value for any cooperative games, not only the DR programs analyzed in this paper.  Also, this method ensures that the ``budget balancing'' constraint is met.  To our knowledge, this constraint has not previously been considered in other research on estimating the Shapley Value using random sampling techniques.  However, its importance is clear in DR programs such as those described here, where a given penalty needs to be distributed in its entirety among participants.

\appendix
\subsection{Proof of Lemma~\ref{lma:neyman_allocation}}
\label{app:A}
The variance of the sample mean of stratum $j$ is
\begin{align*}
	\mathrm{var}(\hat{ \mu}_{j,i})&=\frac{\sigma_{j,i}^2}{N_{j,i}}.
\end{align*}
\noindent
The optimization problem is therefore:
\begin{align*}
\textrm{minimize}\enspace &  \mathrm{var}(T(\rho_{k,j}^i))  \\
\textrm{subject to}\enspace &  \sum_{j=0}^{n-1} N_{j,i} = N.
\end{align*}
Forming the Lagrangian
\begin{align*}
	\mathcal{L} = \frac{1}{n^2} \sum_{j=0}^{n-1}{\frac{\sigma_{j,i}^2}{N_{j,i}}} + \lambda \left(\sum_{j=0}^{n-1} N_{j,i} - N \right).
\end{align*}
Differentiating and setting equal to $0$ results in
\begin{align*}
	N_{j,i} &= N \frac{\sigma_{j,i}}{\sum_{m=0}^{n-1}\sigma_{m,i}}.
\end{align*}
Substituting back yields
\begin{align*}
	\mathrm{var}(T(\rho_{k,j}^i)) 
	&= \frac{1}{N} \left(\frac{1}{n}\sum_{j=0}^{n-1} \sigma_{j,i}\right)\cdot \left( \frac{1}{n}\sum_{m=0}^{n-1} \sigma_{m,i}\right),\\
	\mathrm{var}(\hat{\phi}_i) &= \frac{1}{N} \left(\mathrm{mean}(\sigma_{j,i})\right)^2.
\end{align*}	
\vspace{-0.5cm}
\vspace{-0.2cm}
\subsection{Proof of Equation~\eqref{eqn:var_RS}}
\label{app:B}
A formulation for the Shapley Value is (see equation~\eqref{eqn:shapley_value_1})
\vspace{-0.1cm}
\begin{align*}
	\phi_i =& \frac{1}{n!}\sum_R \rho_{i}(P_{i}^R),
\end{align*}
where $R$ is an ordering of all players and $P_i^R$ is the set of players which precede $i$ in the order $R$.  The statistic $T(\rho_{k,j}^i)$ is calculated by a  random sampling of orderings $R$, i.e., if $N$ random orderings $R$ are analyzed, the statistic is
\begin{align*}
	T(\rho_{k,j}^i) =& \frac{1}{N} \sum_R \rho_i(P_i^R).	
\end{align*}	
Assuming the marginal contributions are uncorrelated, the variance of the estimate is 
\begin{align*}
	\mathrm{var}\left(T(\rho_{k,j}^i)\right) =& \frac{1}{N^2}\sum_R\mathrm{var}\left( \rho_i(P_i^R)\right).
\end{align*}
We define the random variable $\mathcal{J}$ to be a discrete uniform distribution with support $\{0, 1, 2, \ldots, n-1$, and a sample from this distribution is therefore a random strata index, $j$. By the law of total variance and conditioning on $\mathcal{J}$, we then have
\begin{align*}
  \mathrm{var}\left(\rho_i\left(P_i^R\right)\right) &= \mathbb{E}\left[\mathrm{var}\left(\rho_i\left(P_i^R\right) | \mathcal{J}\right)\right]\\&\qquad +\mathrm{var}\left(\mathbb{E}\left[\rho_i\left(P_i^R\right) | \mathcal{J}\right]\right).
\end{align*}
We note that
\vspace{-0.1cm}
\begin{align*}
	 \sigma_{j,i}^2 &= \mathrm{var}\left(\rho_i\left(P_i^R\right) | \mathcal{J}\right) \; \text{and}\; \mu_{j,i} = \mathbb{E}\left[\rho_i\left(P_i^R\right) | \mathcal{J}\right].  
\end{align*}
Therefore,
\vspace{-0.1cm}
\begin{align*}
	\mathrm{var}\left(\rho_i\left(P_i^R\right)\right)&=\mathrm{ mean}(\sigma_{j,i}^2) + \mathrm{var}\left(\mu_{j,i}\right).
\end{align*}
Substituting back gives
\begin{align*}
	\mathrm{var}\left(T(\rho_{k,j}^i)\right) &= \frac{1}{N^2}\sum_R \left[\mathrm{mean}(\sigma_{j,i}^2) + \mathrm{var}\left(\mu_{j,i}\right)\right]\\
	&= \frac{1}{N} \left[\mathrm{mean}(\sigma_{j,i}^2) + \mathrm{var}\left(\mu_{j,i}\right)\right].
\end{align*}
\vspace{-0.5cm}

\bibliographystyle{IEEEtran}
\bibliography{IEEEtranBSTCTL,references}

\end{document}